\documentstyle[12pt,epsf]{article}

\def\DRAFT{0}

\newcommand{\eq}{\begin{equation}}
\newcommand{\en}{\end{equation}}

\newcommand{\eqa}{\begin{eqnarray}}
\newcommand{\ena}{\end{eqnarray}}
\newcommand{\eqan}{\begin{eqnarray*}}
\newcommand{\enan}{\end{eqnarray*}}
\newcommand{\bear}[1]{\begin{array}{#1}}
\newcommand{\enar}{\end{array}}

\newcommand{\lbl}[1]{ \ifnum \DRAFT = 0
                             {\label {#1}}
                      \else  {\makebox[0in]{\raisebox{-2ex}{\tiny #1}
                                            \hspace{-6ex}}}
                             {\label {#1}}
                      \fi}
\newcommand{\rf}[1]{  \ifnum \DRAFT = 0
                             \ref {#1}
                      \else {\ref {#1}}
                            \makebox[0in]{\raisebox{-2ex}{\tiny #1}}
                      \fi}
\newlength{\superigor}
\newcommand{\spazio}[1]{\settowidth{\superigor}{${#1}$}
                        \makebox[\superigor]{} }
\newcommand{\sect}[1]{\setcounter{equation}{0}\section{#1}}

\newcommand{\draft}{ \ifnum \DRAFT =0  {} \else {\\ DRAFT} \fi}

\newcommand{\NP}[1]{Nucl. Phys.\ {\bf #1}\ }

\newcommand{\PR}[1]{Phys. Rev\ {\bf #1}\ }

\def\sqr#1#2{{\vcenter{\hrule height.#2pt
     \hbox{\vrule width.#2pt height#1pt \kern#1pt
        \vrule width.#2pt}
     \hrule height.#2pt}}}

\def\thinspace{\kern .16667em}

\def\Dir{\nabla\kern-2ex\Big{/}}
\def\dslash{\partial\kern-1.5ex\Big{/}}

\def\reali{{\hbox{\s@ l\kern-.5ex R}}}
\def\naturali{{\hbox{\s@ l\kern-.5ex N}}}
\def\interi{{\mathchoice
 {\hbox{Z\kern-1.5mm Z}}
 {\hbox{Z\kern-1.5mm Z}}
 {\hbox{{Z\kern-1.2mm Z}}}
 {\hbox{{Z\kern-1.2mm Z}}}  }}

\def\unity{{\hbox{\s@ 1\kern-.8mm l}}}
\def\uno{{\hbox{ 1\kern-.8mm l}}}
\def\part{\partial}

\def\rd{\sqrt{2}}
\def\um{{1\over2}}
\def\usrd{{1\over\sqrt{2}}}

\def\Llrarr{\Longleftrightarrow}

\def\lrarr{\leftrightarrow}
\def\rarr{\rightarrow}

\def\dag{\dagger}

\def\CA{{\cal A}}
\def\CC{{\cal C}}

\def\CL{{\cal L}}
\def\CM{{\cal M}}

\def\cc{\chi}

\def\dd{\delta}

\def\ee{\epsilon}

\def\ff{\phi}

\def\vf{\varphi}
\def\gg{\gamma}

\def\ll{\lambda}
\def\LL{\Lambda}

\def\pp{\psi}

\def\pb{\bar\psi}

\def\ss{\sigma}
\def\SS{\Sigma}

\newcommand{\mat}[4]{\left(
                     \begin{array}{cc}
                     {#1} & {#2} \\
                     {#3} & {#4}
                     \end{array}
                     \right)
                    }
\newcommand{\vett}[2]{\left(
                      \begin{array}{c}
                     {#1} \\
                     {#2}
                     \end{array}
                     \right)
                    }
\newcommand{\vet}[2]{\left(
                     \begin{array}{cc}
                     {#1} &  {#2}
                     \end{array}
                     \right)
                    }

\begin{document}

\begin{titlepage}

\begin{flushright}
NORDITA-95/3 P\\
January 1994\\
hep-th/yymmddd
\draft
\end{flushright}
\vspace*{0.5cm}

\begin{center}
{\bf
\begin{Large}
{\bf
THE GROSS-NEVEU MODEL ON THE LIGHT CONE
\\}
\end{Large}
}
\vspace*{1.5cm}
          {\large Igor Pesando}
           \footnote{E-mail PESANDO@NBIVAX.NBI.DK, 22105::PESANDO,
                     31890::I\_PESANDO.}
           \footnote{Work supported by the EU grant ERB4001GT930141.}
         \\[.3cm]
          NORDITA\\
          Blegdamsvej 17, DK-2100 Copenhagen \O \\
          Denmark\\
\end{center}
\vspace*{0.7cm}
\begin{abstract}
{ We consider the (massive) Gross-Neveu model using the light cone
quantization where we solve  the constraints explicitly.  We show that
the vacuum is trivial and that the quantization fails when $m=0$.  We
discuss how the running coupling constant emerges as a pure normal ordering
effect in this context and the bound state equation.  }
\end{abstract}
\vfill
\end{titlepage}

\setcounter{footnote}{0}

\def\dmu{\part^{-1}_-}
\sect{Introduction.}

One of the major outstanding problems of the contemporary physics is
to find a way to compute observables in strongly interacting field
theories, of which the theory for the strong interactions $QCD_4$ is
the best examples.  Many methods had been used but for now the most
promising are lattice calculations and light cone field theory: the
latter has received a lot of attention during the last period.  The
advantages of the light cone are that the maximum number of Poincar\'e
generators become kinematical (at $x^+=0$) and that it is very easy to
write a formula for the mass spectrum.  Until now the approach has
been successfully used to solve a variety of 2D problems such as large
$N$ $QCD_2$ with vector matter (\cite{BP}) and adjoint matter
(\cite{Ku}), and applied to $(\ff^4)_{1+1}$ (\cite{PDH}) and $QED_4$
(\cite{KPW}).

A  different approach based on the renormalization group has been
advocated and
is developing by Wilson and collaborators (see for example (\cite{WWHZPG})

In view of a better understanding of the light cone approach we consider the
first non trivial, i.e. non superrenormalizable model in 2D, the Gross-Neveu
model.
The (massive) Gross-Neveu model was previously treated in the infinite
momentum frame (see for example (\cite{Ho})) and in the light cone without
solving explicitly the constraint and assuming the triviality of the vacuum
(\cite{TO}).
Instead we will solve the constraints explicitly and we will prove that the
vacuum is  trivial but this approach  fails when $m=0$, nevertheless
it yields the running coupling constant as a pure normal ordering effect and
the range of stability of the theory in a very simple way.

\sect{Gross-Neveu model in the light cone.}
The lagrangian of the (massive) Gross-Neveu model (\cite{GN}) is given by
\eq
\CL=\pb\cdot(i \stackrel{\lrarr}{\dslash}-m)\pp+{g^2\over N}(\pb\cdot\pp)^2
\en
that can be explicitly written in the light cone as\footnote{
{\bf Conventions.}
$$
x^\pm=x_\mp=\usrd(x^0\pm x^1)~~~~
A^\mu B_\mu=
A_0 B_0 - A_1 B_1=
A_+ B_- + A_+ B_-
{}~~~~
\ee^{01}=-\ee^{+-}=1
$$
$$
\gg_+=\mat{0}{\rd}{0}{0}~~~~
\gg_-=\mat{0}{0}{\rd}{0}~~~~
\gg_0=\mat{0}{1}{1}{0}~~~~
\gg_1=\mat{0}{1}{-1}{0}~~~~
$$
$$
\gg_5=-\gg_0\gg_1=\mat{1}{0}{0}{-1}~~~~
P_{R,L}={1\pm\gg_5\over2}
$$
$$
\pp=\vett{\pp_+}{\pp_-}~~~~
\pb=\vet{\pb_-}{\pb_+}~~~~
 \cc\pb=-\usrd
\mat
{\rd\pb P_R\cc}
{\pb \gg_-\cc}
{\pb \gg_+\cc}
{\rd\pb P_L\cc}
$$
$$
\int_x=\int d^2 x ~~~~
\int_p=\int {d^2 p\over (2\pi)^2}
$$
}

\eq
\CL=i\rd(\pb_+\cdot\stackrel{\lrarr}{\part_+}\pp_+
+\pb_-\cdot\stackrel{\lrarr}{\part_-}\pp_-)
-m(\pb_+\cdot\pp_- +\pb_-\cdot\pp_+)
+{g^2\over N}(\pb_+\cdot\pp_- +\pb_-\cdot\pp_+)^2
\lbl{gn-lag}
\en
where $\pp=(\pp^i)=\pb^*$ with $i=1\dots N$.
As it is usual in the light cone approach primary constraints are
given by the classical equation of motion for the nonpropagating fields
$\pp^i_-$
\eq
-i\rd \part_-\pb_-^i -m\pb_+^i
+2{g^2\over N}(\pb_+\cdot\pp_- +\pb_-\cdot\pp_+)~\pb_+^i
=0
\lbl{eq-mot}
\en
and $\pb^i_-$
\eq
i\rd \part_-\pp_-^i -m\pp_+^i
+2{g^2\over N}(\pb_+\cdot\pp_- +\pb_-\cdot\pp_+)~\pp_+^i
=0
\lbl{eq-mot-b}
\en
Using these constraints we can rewrite the lagrangian (\rf{gn-lag}) as
\eq
\CL'=i\rd\pb_+\cdot\part_+\pp_+
-{m\over2}(\pb_-\cdot\pp_+ +\pb_+\cdot\pp_-)
\en
where $\pp_-$ is to be seen as a functional of $\pp_+$.
{}From the previous effective lagrangian we get the translation generators as
\eqa
{ P}^-
&=&{m\over 2}\int d x^-
{}~\pb_+\cdot\pp_- +\pb_-\cdot\pp_+
\nonumber\\
{ P}^+
&=&i\rd\int dx^- \pb_+\cdot\part_-\pp_+
\lbl{tran-gen}
\ena

These generators are hermitian because we started from a real lagrangian,
i.e. with the explicit $(\pb\cdot\stackrel{\lrarr}{\dslash}\pb)$; would
we not have used such a real lagrangian we would not have got
hermitian generators.

Now we quantize with the standard Dirac brackets
\eq
\{\pp_+^i(x),\pb_+^j(y)\}|_{x^+=y^+}=\usrd \dd^{i j} \dd(x^--y^-)
\lbl{com-rel}
\en
in the light cone box $x^-\in [-L, L]$ where we impose the standard
antiperiodic boundary condition
\eq
\pp^i_+(x^- +2L)=-\pp^i_+(x^-)
{}~~~~
\pb^i_+(x^- +2L)=-\pb^i_+(x^-)
\en
When we expand the operator $\pp_+$ in Schr\"odinger picture in Fourier modes
\eqa
\pp_+^i(x)={1\over \sqrt[4]{2}}\sum_{r\in\interi+\um}\pp_r^i
                             {e^{\pi i r {x\over L}}\over\sqrt{2L}}
\nonumber\\
\pb_+^i(x)={1\over \sqrt[4]{2}}\sum_{r\in\interi+\um}\pb_r^i
                             {e^{-\pi i r {x\over L}}\over\sqrt{2L}}
\ena
eq. (\rf{com-rel}) is equivalent to impose the anticommutation relations
\eq
\{\pp_r^i,\pb_s^j\}=\dd_{r s}\dd^{i j}
\en
With in mind the idea of using a variational approach similar to that used by
Coleman (\cite{Col}) to show that the perturbative vacuum is the true vacuum
(for a certain range of parameters), we introduce the normal order
$N_\CA[\dots]$ defined by
\eq
\pp_r=
\left\{
\bear{l}
 {\mbox{if }~r\in\CC~\mbox{creation operator}}
\\
 {\mbox{if }~r\in\CA~\mbox{annihilation operator}}
\enar
\right.
{}~~~~
\pb_r=
\left\{
\bear{l}
 {\mbox{if }~r\in{\bar\CA}=\CC~\mbox{annihilation operator}}
\\
 {\mbox{if }~r\in{\bar\CC}=\CA~\mbox{creation operator}}
\enar
\right.
\en
where $\CA\cup\CC=\interi+\um$, $\CA\cap\CC=\oslash$.
Since we require that the vacuum $|\CA>$ be C-invariant
\footnote{
We define the charge conjugation as $C\pp_+(x^-)C^{-1}=\pb_+(x^-) \Llrarr
C\pp_rC^{-1}=\pb_{-r}$.
Note that the vacuum $C$-invariance is a consequence of the
request of having a $SO(2N)$ invariant vacuum. This symmetry is
evident when the lagrangian is written in terms of Majorana fermions.
}
 we have to impose $r\in\CA\Llrarr -r\in\CC$.
The choice of the set $\CA$ is equivalent to consider as vacuum the state
\eq
|\CA>\propto
{\prod_{s\in\CA}\prod_{i=1}^N \pp^i_s|0>}
\en
where $|0>$ is the usual free vacuum, defined as $\pp_{-r}|0>=\pb_r|0>=0$ for
$r>0$.

After this introductory stuff we can try to solve the constraints explicitly
and then to write down the explicit form of the translation generators
(\rf{tran-gen}).
Actually, we are not interested in solving  eq.s (\rf{eq-mot-b}) and
(\rf{eq-mot}) but our aim is  to express the operator
$T=\pb_+\cdot\pp_- +\pb_-\cdot\pp_+ $
in function of the propagating fields $\pp_+$ and $\pb_+$.

The formal solution of the constraint (\rf{eq-mot-b}) reads
\eq
\pp^i_{-}=-{i~m\over\rd}\part_-^{-1}\pp^i_+
  +{i\rd g^2\over N}
    \part_-^{-1}[\pp^i_+ (\pb_+\cdot\pp_{-}+\pb_{-}\cdot\pp_+) ]
\lbl{ricor-pp}
\en
and we {\it take } this expression to be the quantum constraint, from
which we derive the quantum constraint for the operator $\pb_-$ by
hermitian conjugation.
{}From this equation we derive the equation for $T$
\eqa
T&=&
-{i m\over\rd} (\pb_+\cdot \part^{-1}_-\pp_+ -\part^{-1}_-\pb_+\cdot\pp_+)
\nonumber\\
&+&{i\rd g^2\over N}[\pb_+\cdot \part^{-1}_-(\pp_+ T)
                     -\part^{-1}_-(T \pb_+)\cdot\pp_+]
\lbl{eq-T}
\ena
The solution of this equation can be given as an expansion in power of
the bilinear $\pb_+(x)\cdot\pp_+(y)$; the explicit solution in the
leading order in $1\over N$ is given in the appendix, here instead we want to
make some comments:
\begin{itemize}
\item{
Since we quantize in a finite box we want to impose antiperiodic boundary
conditions on $\pp_+$ and on $\pp_-$ (obvious if we think to the Dirac
procedure).

This means that the solution (\rf{ricor-pp}) has to "propagate"
the antiperiodic boundary condition from $\pp_+$ to $\pp_-$ and
this in turn requires that we define
properly $\part_-^{-1}$, i.e. in such a way that $\pp_{-}$ satisfies
antiperiodic boundary conditions.
Explicitly we have
\eq
\part^{-1}_-(x)=\sum_r
{e^{\pi i r {x\over L}}\over 2\pi i r}
\en
}

\item{
Notice that there is a possible source of ambiguity: this is due to the
order of $\pp^i_+$ w.r.t. $(\pb_+\pp_{-}+\pb_{-}\pp_+)$ in the second
term of (\rf{ricor-pp}). This ambiguity does not show up to the leading order
as the explicit computation reveals.
}

\item{
What happens to the solution in the limit $m\rarr 0$? As it is clear from the
eq. (\rf{ricor-pp}) the perturbative solution in $g^2$ vanishes. This
is not the right answer because $\pp_-$
has to be antiperiodic. Moreover the hamiltonian $P^-$ also vanishes in this
limit, this means that we cannot quantize the theory on the light cone when
$m=0$.
If we still insist in finding a solution for the $\pp_-$ at $m=0$  we have to
introduce further constraints on the theory and as the $N=1$ case shows, we
have to quantize the values of the charge $\int_{-L}^L dx^-~\pb_+\pp_+$.
}
\end{itemize}
As we are interested in the mass spectrum in the large $N$ limit, we
need not compute the exact solution of (\rf{eq-T}) but we can extract
the leading contributions in $1\over N$.
More precisely we want to compute the mass spectrum of the particles and
of the "mesonic states" hence
we need compute $P^-$ in the approximation
\eqa
P^-&=&N (2L) L_{0}
+\int_{x,x_1,y_1} L_{1}(x;x_1,y_1)
N_\CA[\pb_+(x_1)\cdot \pp_+(y_1)]
\nonumber\\
&~&+{1\over N}\int_{x,x_i,y_i}{L}_2(x;x_1, y_1, x_2, y_2)
N_\CA[\pb_+(x_1)\cdot \pp_+(y_1)\; \pb_+(x_2)\cdot \pp_+(y_2)]
\nonumber\\
&~&+\mbox{higher orders in $\pb\pp$}
\nonumber\\
&=&P^-_{(0)}+P^-_{(1)}+P^-_{(2)}
+\mbox{higher orders in $\pb\pp$}
\lbl{p-0}
\ena
where all the $L$ are of $O(1)$.
The explicit computation yields:
\eqa
P^-&=&
N(2L){m^2 \SS_\CC(0)\over 4}{1\over  g^2 \SS_\CC(0) -1}
-{M^2 L\over 2 \pi}
 \sum_r {1\over r} N_\CA[\pb_r\cdot\pp_r]
\nonumber\\
&~&-{M^2 L\over 8 \pi^2 N}
 \sum_{r_1,r_2,s_1,s_2}
 \dd_{r_1+r_2,s_1+s_2}~g^2(-r_1+s_1)
\cdot
\nonumber\\
&&
 \left({1\over r_1}+{1\over s_1} \right)
 \left({1\over r_2}+{1\over s_2} \right)
  N_\CA[\pb_{r_1}\cdot\pp_{s_1}\pb_{r_2}\cdot\pp_{s_2}]
\nonumber\\
\lbl{p-}
\ena
along with
\eq
P^+=
-N{\pi\over L}~\sum_{r\in\CC} r
-{\pi\over L}
 \sum_r {r} N_\CA[\pb_r\cdot\pp_r]
\lbl{p+}
\en
where we defined the physical mass of the particles $\pb_r|\CA>$
($r\in\CC$)\footnote{This follows from the mass formula
$M^2=2N_\CA[P^+]N_\CA[P^-]$ at the leading order.}
as
\eq
M^2= {m^2\over ( g^2\SS_\CC(0) -1)^2}
\lbl{mass}
\en
and the running coupling constant
\eq
g^2(n)= {g^2\over 1-{g^2}\SS_\CC(n)}
\lbl{running}
\en
where
\eq
\SS_\CC(n)={1\over 2\pi}\sum_{r} {sgn_\CC(r)\over r+n}
\lbl{ll}
\en
and
\eq
sgn_\CC(r)=\left\{\begin{array}{ll}
                   +1 & r\in\CC \\
                   -1 & r\in\CA \end{array}\right.
\en
We notice that when we normal order w.r.t. the free vacuum the running coupling
constant can be written in the thermodynamic limit, i.e. when $L\rarr0$ with
$p={\pi n\over L}=const$ and $p<<{\pi \LL\over L}={\bar \LL}$ as\footnote{
We make use of
$$
\sum_{\um}^{\LL} {1\over r}= \ln {4 e^{\gg_E}~\LL\over \pi} +
O({1\over\LL})
$$
where $\gg_E$ is the Euler constant and $\LL$ is the UV cutoff.
}
\eq
g^2(p)= {g^2\over 1-{g^2}\SS_0(p)}
={2\pi\over \log{p^2\over{\bar \LL}^2e^{-{2\pi\over g^2}}} }
\en
which shows the asymptotic freedom and coincides with the usual result
(\cite{GN}).

Before discussing the range of stability of the theory coupling constant, we
want to point out that the vacuum expectation value of the fermionic
condensate $\pb\cdot\pp(x)$
is proportional to the vacuum energy density because of the
first of (\rf{tran-gen}):
\eq
<\pb\cdot\pp(x)>={P^-_{\mbox{vacuum}}\over 2L {m\over 2}}=-N M \SS_\CC(0)
\lbl{condensate}
\en
which coincides with the formula (112) of (\cite{Ho}) in the substance.
But we are dealing with a composite operator and this needs an independent
renormalization as eq. (\rf{condensate}) shows clearly since it is divergent.
If we are naive and in analogy with the renormalization of $\ff^2$ operator
in a massive scalar theory in the usual lagrangian perturbation approach
we take
$Z_{\pb\pp}=\left.{\part m\over \part M}\right|_{g^2,\LL}$,
which is valid in the zero momentum renormalization scheme, and we get
$<\pb\cdot\pp(x)>_R=-N {M\over g^2(0)}$
but this is wrong because the integrated operator $\int d x^-~\pb\cdot\pp$
is (proportional to) the translation operator $P^-$, which has been
renormalized such that $<P^->=0$, hence the right renormalization is
\eq
<\pb\cdot\pp(x)>_R=0
\lbl{composite}
\en
This agrees with the zero momentum subtraction of the auxiliaty field $\ss$
(\cite{GN}), which is possible in the massive theory.
Explicitely the effective potential reads
\eq
V_{\mbox{eff}}=\ss^2+i\int {d^2k\over (2\pi)^2}\log (k^2- (m-2g\ss)^2)
\en
which can be renormalized with the condition $V_{\mbox{eff}}''(\ss=0)=2$
and since this means that ${\ss\over g}=<\pb\cdot\pp>=0$
we recover eq. (\rf{composite}).

\sect{The vacuum of the theory.}
Now we want to prove that the vacuum is trivial in a certain range of
$g^2$ and that outside this range the vacuum is not among  the test
states $|\CA>$.

The simplest way of seeing this is to consider
\eq
<\CA| N_{0}[P^-]|\CA>=P^-_{(0)}(\CA)-P^-_{(0)}(0)
\en
This quantity has to be greater than zero because the perturbative
vacuum $|0>$ be the true vacuum.

{}From the plot of the function $P^-_{(0)}(\ll)\propto {\ll\over g^2\ll-1}$
($\ll=\SS(0)$) and from the bound
$|\SS_\CC(0)|\le \SS_0(0)={1\over \pi}\sum_{\um}^\LL {1\over r}=\ll_{max}$
we see that two situations can arise.

\begin{center}
\leavevmode
\epsfbox{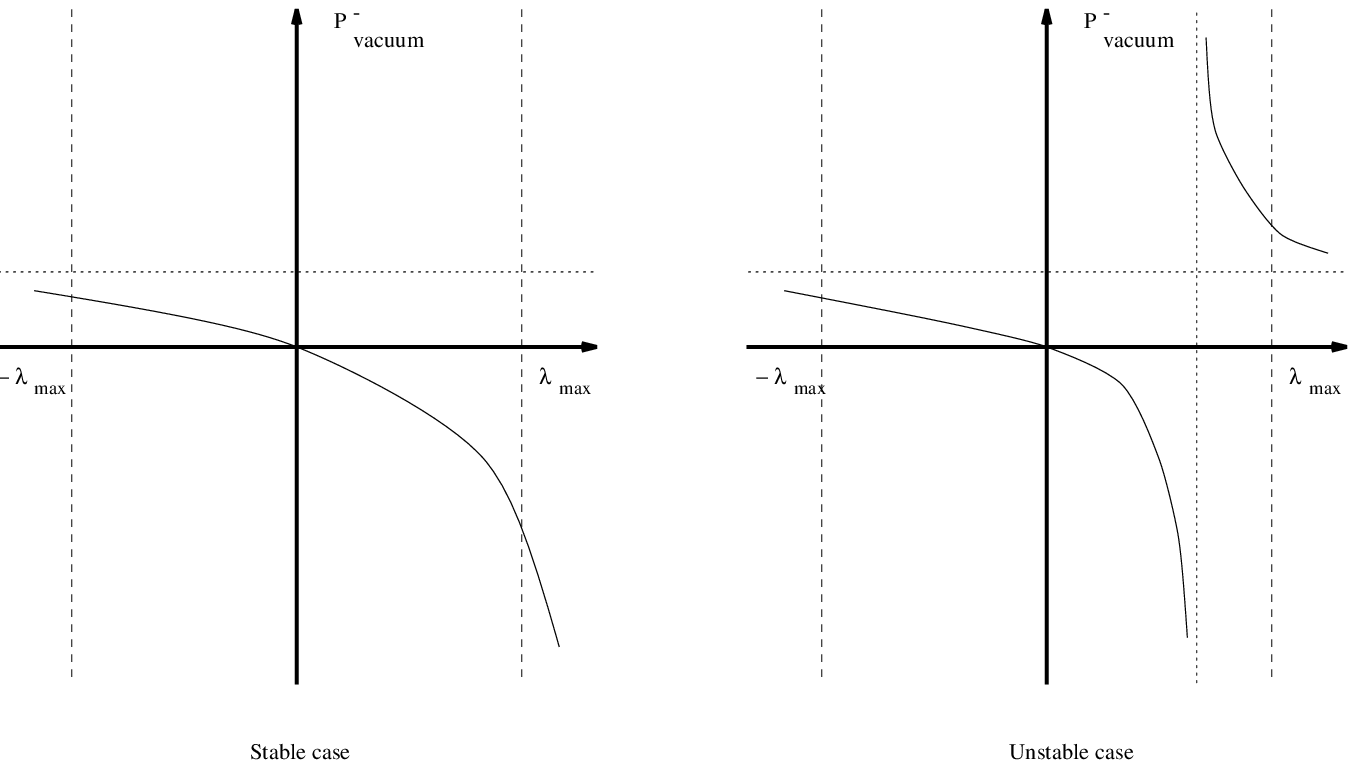}
\end{center}

When
\eq {\pi\over g^2}> \pi\ll_{max} =\sum_{\um}^\LL {1\over r}
\Llrarr {\pi\over g^2(0)}> 0
\lbl{bound1}
\en
the minimum of the vacuum energy is achieved for the perturbative
vacuum while when eq. (\rf{bound1}) is not satisfied, the free vacuum
$|0>$ is not the minimum energy eigenstate of the theory, i.e. it is
NOT the true vacuum.
Now we can ask whether one of the $|\CA>$ states is the true vacuum and
whether the theory is meaningful since its ground state seems to have unbounded
negative energy.

In order to answer to this question and to confirm the previous
coupling constant range we look at the mesonic spectrum and we check for
tachyonic states.
We define the following $U(N)$ invariant operators
\eq
M^\dag_{r s}={1\over\sqrt{N}} \pb_r\cdot\pp_s
{}~~~~
M_{r s}={1\over\sqrt{N}} \pb_s\cdot\pp_r
{}~~~~
r\in\CC, s\in\CA
\en
which satisfy the commutation relations
\eq
[M_{r_1 s_1},M^\dag_{r_2 s_2}]=\dd_{r_1 r_2}\dd_{s_1 s_2}
+O\left({1\over\sqrt{N}}\right)
\lbl{com-rel}
\en
and the look for the mesonic mass spectrum.
We need not the other $U(N)$ invariant operators $\pb_r\cdot\pp_s$ with either
$r,s\in\CA$ or $r,s\in\CC$ because they annihilate the vacuum $|\CA>$ and can
be always brought to act on the vacuum.

With the help of the $M^\dag$ and $M$ we can rewrite the leading order
contribution of $P^-$ when acting on states created by $M^\dag$ as
\eqa
N_{\CA}[P^-]&=&
{M^2 L\over 2\pi} \sum_{r_2,r_2\in\CC~s_1,s_2\in\CA}
M^\dag_{r_1 s_1}
\biggl\{
\left({1\over s_1}-{1\over r_1}\right) \dd_{r_1 r_2} \dd_{s_1 s_2}
\nonumber\\
&-&{g^2(s_1-r_1)\over 2\pi}  \dd_{s_1-r_1,s_2-r_2}
\left({1\over s_1}+{1\over r_1}\right)
\left({1\over s_2}+{1\over r_2}\right)
\biggr\}
 M_{r_2 s_2}
\lbl{ham}
\ena

It is possible to write $P^-$ as in (\rf{ham}) because we are only considering
states generated by $M^\dag$ acting on the vacuum $|\CA>$, which implies that
we only need considering the commutation properties of $P^-$ with the $M$
operators.
In particular we find
\eq
[N_{\CA}[P^-_{(1)}], M^\dag_{r s}]=
{M^2 L\over 2\pi}
\left({1\over s}-{1\over r}\right)
M^\dag_{r s}
\en
which allows us to write
$P_{(1)}^-={M^2 L\over 2\pi} \sum_{s\in\CC,r\in\CA}
\left({1\over s}-{1\over r}\right)
M^\dag_{r s}M_{r s}
$ in virtue of (\rf{com-rel}).
As far as $P^-_{(2)}$ is concerned we are dropping the non leading
contributions.

Now our task is to diagonalize the hamiltonian (\rf{ham})
\eq
N_{\CA}[P^-]={M^2 L\over 2\pi} \sum M^\dag_{r_1 s_1}
\CM_{r_1 s_1, r_2 s_2}M_{r_2 s_2}
\lbl{ham-1}
\en
and to find its eigenvalues.

To this purpose we
notice that the matrix $\CM_{r_1 s_1, r_2 s_2}$ can be written in a block form
where the blocks are characterized by a natural $R=s-r$:
\eqa
\CM(R)_{s_1, s_2}=
\left({1\over s_1}+{1\over R-s_1}\right) \dd_{s_1 s_2}
-{g^2(R)\over 2\pi}
\left({1\over s_1}-{1\over R-s_1}\right)
\left({1\over s_2}-{1\over R-s_2}\right)
{}
\nonumber\\
{}\spazio{\CM(R)_{s_1, s_2}=\left({1\over s_1}+{1\over R-s_1}\right)
          \dd_{s_1 s_2}-{g^2(R)\over 2\pi} }
s_1,s_2\in\CC \mbox{ and } R-s_1,R-s_2\in\CA
\ena
The meaning of the integer $R=s-r={L\over\pi}P^+$ clearly results from
\eq
[N_{\CA}[P^+], M^\dag_{r s}]={\pi\over L} R M^\dag_{r s}
\lbl{p+1}
\en

We can now write the formula for the eigenvalues for the matrices $\CM(R)$:
\eqa
&\det&\left( \CM(R)-\mu \uno\right)
=
\nonumber\\
&&=
\prod_{s\in\CC,s-R\in\CA}\left({R\over s(R-s)}-\mu\right)\cdot
\nonumber\\
&\det&
\biggl|\biggl| \dd_{s_1 s_2}
-{g^2(R)\over 2\pi}
{2s_1 -R\over\sqrt{s_1(R-s_1)}\sqrt{R-\mu s_1(R-s_1)}}
{2s_2 -R\over\sqrt{s_2(R-s_2)}\sqrt{R-\mu s_2(R-s_2)}}
\biggr|\biggr|
\nonumber\\
&&=
\prod_{s\in\CC,s-R\in\CA}\left({R\over s(R-s)}-\mu\right)
\left(1
-{g^2(R)\over 2\pi}
\sum_{s\in\CC,s-R\in\CA}
{(2s -R)^2\over s(R-s)(R-\mu s(R-s))}
\right)
\nonumber\\
{}~
\ena
where we used the fact the a matrix $\parallel \dd_{i j}+ a_i a_j\parallel$
has all but one eigenvalues equal to $1$
If we now exploit the symmetry $s\lrarr R-s$ of the C-invariant test
vacua (where $s\in\CC \Llrarr -s\in\CA$ has as the consequence that
if $s\in\CC$ then $-r=R-s\in\CC$) we get the eigenvalues
\eq
\mu_s={R\over s(R-s)}
{}~~~~
s\in\CC, s\le R-s
\lbl{eig-1}
\en
and the equation (for $R\ge 2$)
\eq
{2\pi\over g^2(R)}=
{2}
\sum_{s\in\CC, s<R-s}
\left(2s -R\over s(R-s)\right)^2{1\over {R\over s(R-s)}-\mu }
\lbl{eig-2}
\en

It is now immediate to see that none of the non trivial $|\CA>$ can be
the true vacuum since from (\rf{eig-1}) we find that there are tachyonic
mesons with mass $M^2_{\mbox{tac}}=-M^2 {(s-r)^2\over s r}$ when
$r,s>0$ or $r,s<0$.

Let us now consider the case of the trivial vacuum.
In  eq. (\rf{eig-1}) the admissible values of $s$ are now
$
s=\um,\dots \left[{R\over 2}\right]+\um
$ and
the eigenvectors associated to these eigenvalues are
\eq
|\mu_s(R)>={\dd_{s-r,R}\over\rd}(M_{r s}^\dag-M_{-s,-r}^\dag)|0>
\en
and describe  C-invariant states made of free particles.
Similarly the summation condition in eq. (\rf{eig-2}) becomes
$0<s<\left[{R\over 2}\right]$.
{}From the graph of the r.h.s. of (\rf{eig-2}) as a function of $\mu$
and from ${\part\over\part\mu}\left({2\pi\over g^2(R)}\right)<0$
it is clear that this equation always has $\left[{R\over 2}\right]$ solutions
and that it only has one negative solution when
\eq
0<{2\pi\over g^2(R)}<
{2\over R}
\sum_{0<s<[R/2]}{(2s -R)^2\over s(R-s)}
\en
The only way to avoid to have mesonic tachyons is to impose
\eq
{2\pi\over g^2}\ge
2\sum_\um^{\LL} {1\over r}
+(\sum_{\LL+1}^{\LL+R} {1\over r}-\sum_{\LL-R+1}^{\LL} {1\over r})
+2 ({1\over R}-2)
{}~~~ \forall R\ge2
\lbl{bound2}
\en
which implies in the limit $1,R<<\LL$
\eq
{2\pi\over g^2}\ge
2\sum_\um^{\LL} {1\over r}-3-O\left({R^2\over\LL^2}\right)
\en
that essentially coincides with (\rf{bound1}) a part from the constant
$-3$:
it is probable that some inconsistencies (tachyonic poles) show up in the range
$0\ge{2\pi\over g^2(0)}>-3$ when we consider the correlation functions of
some (composite) operators.

This relation will turn out to be fundamental in order to prove the Lorentz
invariance of the mesonic spectrum in the thermodynamic limit, since
the continuum version of eq. (\rf{eig-1}) is the consistency equation
for the bound state wave function.

\sect{The 't Hooft equation for the mesonic spectrum.}

The 't Hooft equation is a light cone form for the Bethe-Salpeter equation,
the aim of which is to find the spectrum of the bound states.
This can be easily obtained from eq. (\rf{ham-1}), (\rf{eig-2}) and
(\rf{p+1}): the
spectrum is given by
\eq
M^2_{\mbox{meson}}=2{\pi\over L} R~ {M^2 L\over2\pi}\mu(R)=M^2 R\mu(R)
\lbl{m-box}
\en
where $\mu(R)$ is any of the solutions of the eq. (\rf{eig-2}).
But actually we are interested in the spectrum in the thermodynamic
limit, i.e. when $R\rarr\infty$ with ${R\over L}=const$ as
$L\rarr\infty$.
This means we are interested in the solutions of (\rf{eig-2}) that
scale as $1\over R$.
If we set $\mu={\ss\over R}$ and $s=x R$, we can rewrite (\rf{eig-2})
in thermodynamic limit as
\eq
{2\pi\over g^2(R)}=2\int^{\um-{1\over 2R}}_{{1\over 2 R}} dx~ {(1-2
x)^2\over x(1-x)}{1\over 1- \ss x(1-x)}
\lbl{cons-eq}
\en
where both sides are divergent but their divergences cancel exactly,
and what is left it is independent of $R$ and yields a Lorentz
invariant result for the meson mass.
This is analogous to what happens in $QCD_2$ when the 't Hooft IR
cutoff $\ll$ (\cite{tH}) is used.
Moreover this sheds light on the nature of the bound state wave function
renormalization as done in (\cite{Cav}): it is an expression of the
fact that the IR divergences in the thermodynamic limit can be seen as
UV divergences in the box.

We can also get the 't Hooft equation in a more conventional way; we
define the mesonic state as
\eq
|\vf,R>=\sum_s \vf_R(s) M^\dag_{s-R,s}|0>
\en
and we get the 't Hooft equation requiring it to be the eigenstate of
the mass equation $M^2=2 P^+ P^-$, explicitly
\eqa
{M^2_{\mbox{meson}}\over M^2}\vf(x)&=&
\left({1\over x}+{1\over 1-x}\right)\vf(x)
\nonumber\\
&&-{g^2(R)\over2\pi}\left({1\over x}-{1\over 1-x}\right)
\int^{1-{1\over 2R}}_{1\over 2R}dy~\left({1\over y}-{1\over 1-y}\right)\vf(y)
\ena
The solution of this equation
\eq
\vf(x)={1-2x\over x(1-x)}
       {1\over {1\over x(1-x)}-{M^2_{\mbox{meson}}\over M^2} }
       K(R)
\en
does not seem to have a Lorentz covariant spectrum, but the
consistency equation for the factor $K(R)={g^2(R)\over2\pi}
\int^{1-{1\over 2R}}_{1\over 2R}dy~\left({1\over y}-{1\over 1-y}\right)\vf(y)$
is again (\rf{cons-eq}).

\sect{Conclusion.}

In this work we considered the Gross-Neveu model and we
quantized it using the light-cone approach in order to  understand better how
renormalization and dynamical symmetry breaking come into the play.
We discovered that we cannot describe the massless Gross-Neveu model and hence
we cannot describe the ``dynamical symmetry breaking''.
We found also that
in the massive case the running of the coupling constant is only due to normal
ordering effects but we have not an explanation for this phenomenon.

A way to describe both the massless and the massive case is to consider the
Yukawa lagrangian
$$
\CL=
\pb\cdot (i\stackrel{\lrarr}{\part\kern-.5em \Big{/}}-m)\pp
+\um (\part\ff)^2-\um \mu \ff^2
+g\mu\pb\cdot\pp~\ff
$$
and let $\mu\rarr\infty$.

Another interesting point would be to discuss the kinks in this formalism.
All these things are left for future work.

\vspace*{1cm}
{\bf Acknowledgement.}
I thank P. Di Vecchia for useful discussions.

\section{Appendix 1.}
\appendix
In view of eq. (\rf{p-0})
we want to compute the following contributions
\eqa
T(x)=(\pb_{-}\pp_+&+&\pb_{+}\pp_-)(x)=
NT_{0}(x)
+ \int_{x_1,y_1} T_{1}(x;x_1,y_1) N_\CA[\pb_+(x_1)\cdot \pp_+(y_1)]
+\dots
\nonumber\\
&=&\sum_{m=0}^{\infty}N^{1-m}\int_{(x_i,y_i)_{1\le i\le m}}
T_{m}(x;(x_i, y_i)_{1\le i\le m})
N_\CA[\prod_{i=1}^{m}\pb_+(x_i)\cdot \pp_+(y_i)\; ]
\nonumber\\
\;
\lbl{coeff-def}
\ena
where $T=O(1)$ and we neglect the non leading contribution to the $T$.
Since T.s are the terms with the highest $N$ power,
we only have to consider in (\rf{eq-T}) the contributions that come
from the normal order of the pair $\pb_+(x)\cdot\pp_+(y)$ where the
 $U(N)$ indices are contracted because the other are suppressed in $1\over N$.
With this observation in mind we can set up a set of relations
for the coefficients in (\rf{coeff-def}).
The actual  relations are
\eq
T_{0}(x)=
 -{g^2}\int_y K(x-y)T_{0}(y)
 -{m\over 2}\int_y K(x-y)
\lbl{rec-rel-l0}
\en
and
\eqa
T_{1}(x;x_1,y_1)
&=&
 -{g^2}\int_y K(x-y)T_{1}(y;x_1,y_1)
\nonumber\\
 &&-i\rd \dmu(x_1-y_1)
 \left[\dd(x_1-x) \left( {m\over 2}-g^2 T_{0}(y_1)\right)
      +\dd(y_1-x) \left( {m\over 2}-g^2 T_{0}(x_1)\right)\right]
\nonumber\\~~
\lbl{rec-rel-l1}
\ena
and more generally (notice that it is not necessary to explicitly
symmetrize w.r.t. the $x_i$ and $y_i$ indices since the product
$N_\CA[\prod_{i=1}^{m}\pb_+(x_i)\cdot \pp_+(y_i)\; ]$ is already symmetric)
\eqa
 T_{m}(x;(x_i,y_i)_{1\le i\le m})&=&
 -{g^2}\int_y K(x-y)T_{m}(y;(x_i,y_i)_{1\le i\le m})
\nonumber\\
&&+g^2\dmu(x_m-y_m)
      \left(\dd(x_m-x) T_{m-1}(y_m;(y_j,x_j)_{1\le j\le m-1})
\right.
\nonumber\\
&&\left.
           +\dd(y_m-x) T_{m-1}(x_m;(y_j,x_j)_{1\le j\le m-1})
      \right)
\lbl{rec-rel-l2}
\ena
where we defined
\eq
K(x-y)={\rd\over N}~i\dmu(x-y)
 (\underbrace{\pb_+(x)\cdot\pp_+(y)}
 -\underbrace{\pb_+(y)\cdot\pp_+(x)})
\lbl{kernel}
\en
A simple computation yields
\eq
{\rd\over N}\underbrace{\pb_+(x)\cdot\pp_+(y)}=
 \sum_{r\in\CC} {e^{\pi i r {y-x\over L}}\over 2L}
\en
and
\eq
K(x-y)=\sum_n {e^{\pi i n {x-y\over L}}\over2L}
         \sum_s{sgn_{\CC}(s)\over 2\pi(s+n)}
\en
It is now easy to get from (\rf{rec-rel-l0}) and (\rf{rec-rel-l1})
eq. (\rf{p-}).

The solution for eq. (\rf{rec-rel-l2}) reads
\eq
 T_{m}(x;(x_i,y_i)_{1\le i\le m})=
 2^{m}\prod_{1\le i\le m} \dmu(x_i-y_i)
 g^2_{x(x_m}g^2_{y_m)(x_{m-1}}\dots g^2_{y_2)(x_1}T_{y_1)}^{(0)}
\en
where $g^2_{xy}=g^2 (\uno-g^2 K)^{-1}_{xy}$.
This formula shows that all the terms can be expressed in function of
the running coupling constant at different momenta.

\end{document}